\documentclass[12pt]{article}
\usepackage[utf8]{inputenc}
\usepackage{amsfonts}
\usepackage{latexsym}
\usepackage{amsmath,amssymb}
\usepackage{graphicx}
\usepackage{verbatim}
\usepackage{setspace}
\usepackage{color}
\usepackage{mathdots}
\usepackage{yhmath}
\usepackage{cancel}
\usepackage{color}
\usepackage{mathrsfs}  
\usepackage{array}
\usepackage{multirow}
\usepackage{amssymb}
\usepackage{tabularx}
\usepackage{hyperref}
\usepackage{booktabs}
\numberwithin{figure}{section}

\usepackage[letterpaper,hmargin=0.8in,vmargin=1in]{geometry}
\numberwithin{equation}{section}

\begin{document}
\unitlength = 1mm
\ \\
\vskip1cm
\begin{center}

{ \LARGE {\textsc{Marginal Operators from Celestial Diamonds}}}

\vspace{0.8cm}
Michael Imseis${^{\ast,\dagger}}$, Sruthi A. Narayanan$^\dagger$ and A.W. Peet${^\ast}$

\vspace{1cm}

{\it ${}^\ast$University of Toronto Department of Physics,\\
McLennan Physical Laboratories, 60 St George St, Toronto, ON M5S 1A7, Canada}

{\it ${}^\dagger$Perimeter Institute for Theoretical Physics,\\
31 Caroline Street North, Waterloo, ON N2L 2Y5, Canada }

\begin{abstract}
For a given conformal field theory (CFT), one can deform it via the addition of a marginal operator to the spectrum. In two dimensions, when the added operator has conformal weights $h=\bar{h}=1$, conformal symmetry is not broken and the resulting theory is a distinct CFT. Studying such marginal operators for celestial CFTs allows for a geometric understanding of the space of allowed boundary theories dual to quantum field theories (QFT) in bulk asymptotically flat spacetimes. In traditional holographic examples, a marginal deformation on the boundary corresponds to a vacuum transition in the bulk theory. We affirm this in celestial CFTs which requires a general definition of marginal operators as composite celestial operators via pairs that live at distinct corners of celestial memory and Goldstone diamonds.
 \end{abstract}
\vspace{0.5cm}

\vspace{1.0cm}

\end{center}

\pagestyle{empty}
\pagestyle{plain}
\newpage
\tableofcontents

\pagenumbering{arabic}

\section{Introduction}

Celestial (or flat-space) holography posits a duality between a theory of quantum gravity in asymptotically flat spacetime and a lower dimensional ``celestial" conformal field theory (CFT) that lives on a cut of the null boundary~\cite{Strominger:2017zoo}. Within this formulation, bulk scattering amplitudes are recast as conformal correlators via an explicit map~\cite{Pasterski:2016qvg}. To date, translating the properties of scattering amplitudes to the boundary has given us data about the celestial CFT including insight into the spectrum and operator product expansion (OPE) coefficients. See~\cite{Raclariu:2021zjz, Pasterski:2021rjz} for detailed reviews.

An original goal of the celestial holography program was to exploit this duality and use the boundary CFT to uncover aspects of a bulk theory of quantum gravity. However, it is still an open question as to what the concrete definition of a celestial CFT is. To this end, it is of current interest to find an intrinsic definition of a celestial CFT which is somewhat agnostic to the bulk data. Progress has been made in this direction by considering CFTs that share characteristics with Liouville theories~\cite{Stieberger:2022zyk} or that come equipped with a Wess-Zumino-Witten sector~\cite{Melton:2022fsf, Melton:2023lnz} and successfully reproducing maximal helicity violating (MHV) amplitudes in a bulk theory of Yang-Mills.\footnote{Whenever we discuss a bulk theory of Yang-Mills in the context of celestial CFT, we are implicitly assuming a weak coupling to gravity. Usually, one considers such a theory in a limit where gravity is negligible.} 

A complementary approach is to understand the space of all possible celestial CFTs. The hope is that determining a set of candidates might shed light on the ingredients necessary for an intrinsic definition of the boundary. This is aided by the existence of a beautiful geometric interpretation whereby the space of CFTs can be described by a manifold known as the conformal manifold. Points on the manifold correspond to different CFTs in the same family and one can deform from one to another by the addition of a marginal operator into the CFT spectrum. This manifold is equipped with a metric, connections and curvature that can be derived from the OPEs of the marginal operators~\cite{Kutasov:1988xb}. It is expected that for a holographic theory, parallel transport around the manifold is dual to vacuum transitions in the corresponding bulk theory. This has been studied and confirmed for certain super-conformal field theories~\cite{Green:2010da,Balthazar:2022hzb} but has not been proven for general holographic correspondences.

In~\cite{Kapec:2022axw} this was explored in the context of the non-linear sigma model (NLSM) for celestial CFT. The marginal operators in this case corresponded to  shadow transformed soft scalars in the bulk. The advantage to studying this in the context of a NLSM is that the connection between the conformal manifold and degenerate vacua in the bulk is immediately clear. In this case, parallel transporting in the celestial conformal manifold was equivalent to a vacuum transition in the bulk theory, as conjectured for holographic theories and the geometry of the manifold mirrored that of the target space manifold for the bulk theory. However, the simplicity of this relationship was largely due to the fact that the fundamental fields were scalars which means, from the boundary point of view, they only needed to have the correct conformal dimension in order to be interpreted as marginal operators. The analogous story in gauge theory and gravity turns out to be more complex due to the presence of non-zero spin.

\begin{figure}[thb!]
\begin{center}
\includegraphics[scale=.5]{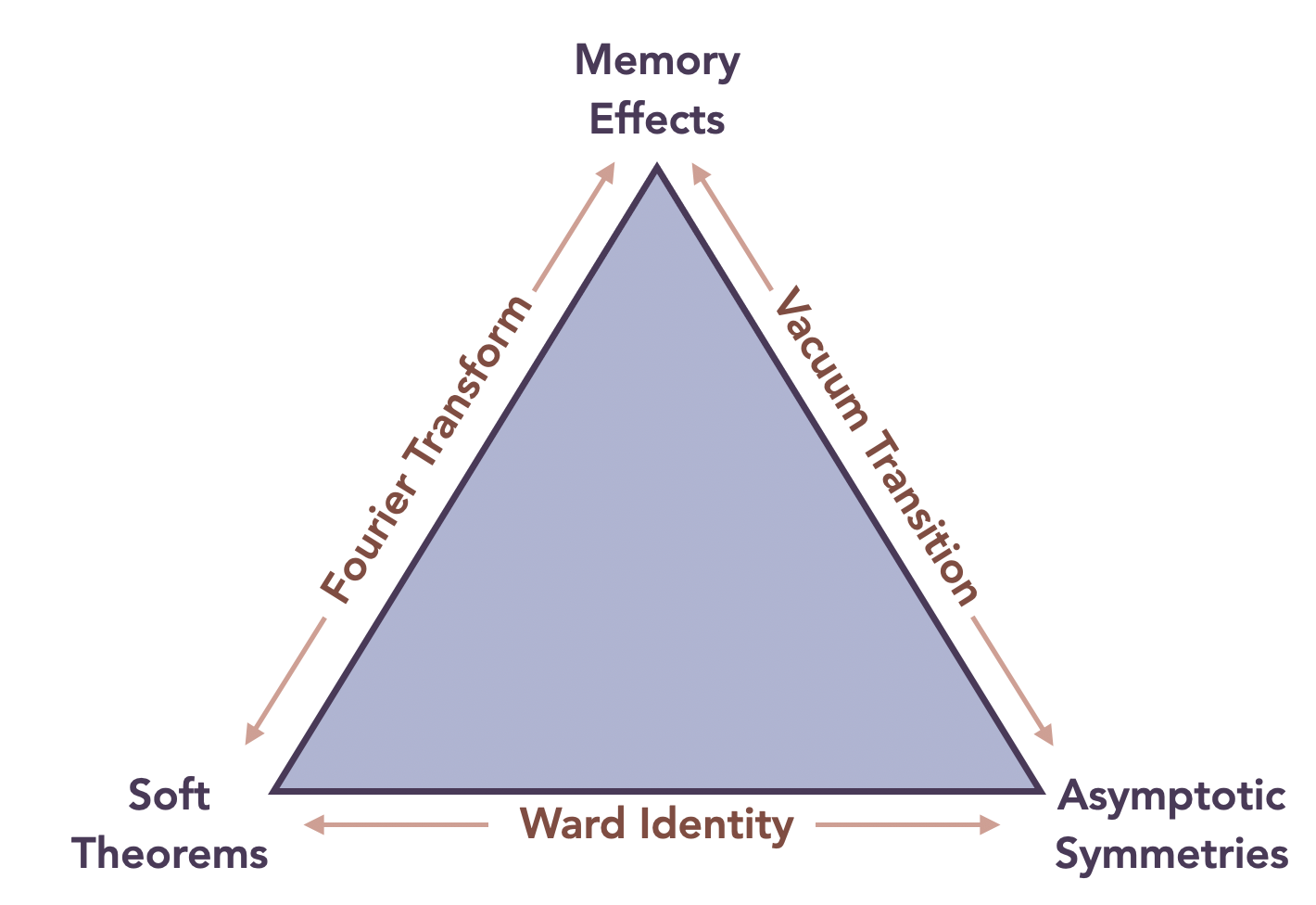}
\caption{The infrared triangle for asymptotic symmetries, as in~\cite{Strominger:2017zoo}. We will concern ourselves with the side that relates memory effects to asymptotic symmetries via vacuum transitions.}
\label{Fig:ir}
\end{center}
\end{figure}
It has long been established that asymptotic symmetries, which are symmetries that preserve the asymptotic falloffs of the metric, can be interpreted as vacuum transitions in the bulk theory. In particular, as seen in Figure~\ref{Fig:ir}, the relationship between asymptotic symmetries and vacuum transitions involves connecting the symmetries to memory effects. For example, if we consider supertranslations which are the asymptotic symmetries corresponding to the leading soft graviton theorem, they act by angle dependent translations along null infinity.\footnote{In fact, supertranslations appear as isometries of null hypersurfaces, viewed as Carrollian manifolds, even at finite distance. See \cite{Ciambelli:2025unn} for a review.} The passage of soft radiation, i.e soft gravitons, will affect measurements made at null infinity which is the definition of leading gravitational memory~\cite{Strominger:2014pwa}. The shift in the measurement is precisely the effect of a supertranslation. The interpretation as a vacuum transition can be seen at the level of the metric. A supertranslation shifts you to a distinct metric that has the same asymptotic falloffs which can be thought of as a distinct gravitational vacuum characterized by the addition of a soft graviton. 

The appearance of infinitely many degenerate vacua in an asymptotically flat spacetime should be somehow consistent with the description of a conformal manifold for the dual CFT. In~\cite{Kapec:2022hih} such a connection was made in a story that, in many aspects, mirrored that of the NLSM. Naively, it seemed as though these operators did not satisfy the conditions to be marginal. They were thereby referred to as ``quasi-marginal" in~\cite{Narayanan:2024qgb}, where the author constructed marginal operators in gauge theory via light transforms as opposed to the shadow transforms considered in~\cite{Kapec:2022hih}. In this work we will comment on the consistency of these results in the framework of marginal deformations.

In gravity, this question becomes more interesting because the construction of marginal operators appears to be more complex that just a non-local transformation of a soft operator. One can see this by considering the Casimirs of the conformal group which are given in terms of spin and dimension as~\cite{2018JHEP...11..102K}
\begin{eqnarray}
C_2(\Delta,J) & = & \Delta(\Delta-2)+J^2 = 2(h(h-1)+\bar{h}(\bar{h}-1))\cr
C_4(\Delta,J) & = & (\Delta-1)^2J^2 = (h+\bar{h}-1)^2(h-\bar{h})^2.
\end{eqnarray}
In 2D, a marginal operator has vanishing $C_2$ and $C_4$. If we wish to identify marginal operators as transforms of primaries, then the transformation must preserve the Casimirs. Therefore, the initial operator must also have vanishing $C_2$ and $C_{4}$. Solving the quadratic equations tells us that 
\begin{equation} \label{1.2}
\Delta_{\pm} = 1\pm\sqrt{1-J^2} \ \ \mbox{and either} \ \ \Delta=1 \ \mbox{or} \ J=0.
\end{equation}
It is then easy to see that for scalars we need to consider an initial operator with $\Delta = 2,0$ as explored in~\cite{Kapec:2022axw}. Similarly, for gluons, $J=\pm 1$, we need $\Delta=1$ as considered in~\cite{Narayanan:2024qgb}. The Casimir  preserving transformations that result in the marginal operators of~\cite{Kapec:2022axw} and~\cite{Narayanan:2024qgb} are the light and shadow transforms which are explicitly given by
\begin{eqnarray}
\mathcal{L}\left[\mathcal{O}_{h,\bar{h}}\right](z,\bar{z}) & = &\int_{\mathbb{R}}\frac{dw}{2\pi i}\frac{1}{(w-z)^{2-2h}}\mathcal{O}_{h,\bar{h}}(w,\bar{w})\cr
\bar{\mathcal{L}}\left[\mathcal{O}_{h,\bar{h}}\right](z,\bar{z}) & = &\int_{\mathbb{R}}\frac{d\bar{w}}{2\pi i}\frac{1}{(\bar{w}-\bar{z})^{2-2\bar{h}}}\mathcal{O}_{h,\bar{h}}(w,\bar{w})\cr
\mathcal{S}\left[\mathcal{O}_{h,\bar{h}}\right](z,\bar{z}) & = &\int d^2w\frac{1}{(w-z)^{2-2h}(\bar{w}-\bar{z})^{2-2\bar{h}}}\mathcal{O}_{h,\bar{h}}(w,\bar{w})
\end{eqnarray}
when acting on a CFT operator of weights $(h,\bar{h})$. However, as soon as we consider $|J|>1$, i.e gravity, \eqref{1.2} can no longer be satisfied. This thereby necessitates a different construction of marginal operators for gravity which we explore in this paper.

In this work we attempt to reconcile the notion of vacuum transitions with exactly marginal operators in gravity and gauge theory, through the lens of the celestial diamonds studied in~\cite{Pasterski:2021fjn,Pasterski:2021dqe} and multi-particle operators studied in~\cite{Guevara:2024ixn}. We begin in section~\ref{sec:celsym} with a review of the symmetries of celestial CFTs. We will collect some important results from the existing literature with regards to the $w$-algebra and translations. In section~\ref{sec:celdiam} we review the notion of primary descendants in celestial CFT and how to package them within the diamond structure for conformal multiplets. In section~\ref{sec:marginal} we provide the necessary background for marginal operators and discuss a general construction in a theory with particles of arbitrary spin. In section~\ref{sec:vacuum} we discuss the implications of this construction for the boundary conformal manifold and comment on its relationship with the bulk vacuum manifold. We conclude in section~\ref{sec:future} with a discussion of some open questions and future directions.

\section{Celestial symmetries}\label{sec:celsym}
In this section we begin with a discussion of symmetries in the context of celestial CFTs for two reasons. One is that the physical implications of the marginal operators have a nice interpretation in terms of symmetries and the other is that symmetries in celestial CFTs are slightly different from standard CFT. 

The group of asymptotic symmetries is the Bondi-Metzner-Sachs (BMS) group, which is the semi-direct product of supertranslations with the Lorentz group. It is also known that the Lorentz subgroup in four dimensions is manifest as the global conformal group in two dimensions~\cite{Pasterski:2016qvg,Pasterski:2017kqt}. The relationship between these two symmetry groups forms the basis for flat space holography. Massless particles\footnote{In this work we are only concerned with massless particles since all soft particles are massless but there is also a way to realize massive particles as operators in celestial CFTs.} in the bulk are understood as operators in the boundary celestial CFT via a Mellin transform in energy
\begin{equation}\label{eq:mellin}
\mathcal{O}_{\Delta,J} \equiv \int_0^\infty d\omega \omega^{\Delta-1}\Phi_{\mu_1\cdots\mu_J}^\pm
\end{equation}
where $\Phi^{\pm}_{\mu_1\cdots\mu_J}$ corresponds to the momentum space wavefunction of a spin-$J$ particle and $\Delta$ is the conformal dimension of the corresponding operator. Transforming fields that correspond to soft insertions in the $S$-matrix give a special class of boundary operators with integer conformal dimension that form the spectrum of the celestial CFT~\cite{Freidel:2022skz,Cotler:2023qwh}. In particular, transforming soft bulk fields give operators whose insertions into celestial conformal correlators generate the conformally soft theorems~\cite{Pate:2019mfs,Puhm:2019zbl}. In gravity, the leading soft graviton corresponds to a supertranslation current on the boundary and the subleading soft graviton corresponds to the generator of superrotations. The latter is especially important since its shadow can be identified as the stress tensor in celestial CFT~\cite{Kapec:2014opa,Kapec:2016jld,Kapec:2017gsg}. 

The generators of the global conformal group are given by the modes of this stress tensor $L_n,\bar{L}_n$ for $n=0,\pm 1$ and they respect the following algebra
\begin{equation}
[L_m,L_n] = (m-n)L_{m+n}, \ \ [\bar{L}_m,\bar{L}_n] = (m-n)\bar{L}_{m+n}.
\end{equation}
As previously mentioned, the BMS group contains these generators, along with the generators of supertranslations, which are captured by a conserved current $P(z,\bar{z})$. The generators of supertranslations are the modes $P_{m,n}$ of this current and their action on celestial primary operators takes the following form
\begin{equation}
\left[P_{m,n}, \mathcal{O}_{h,\bar{h}}(z,\bar{z})\right] = \frac{1}{2}z^{m+\frac{1}{2}}\bar{z}^{n+\frac{1}{2}}\mathcal{O}_{h+\frac{1}{2},\bar{h}+\frac{1}{2}}
\end{equation}
where we have chosen to label the operator by its left and right conformal weights which are related to the dimension and spin by $(h,\bar{h}) = \left(\frac{\Delta+J}{2},\frac{\Delta-J}{2}\right)$. Of particular interest to us is the special case when $m=n=-\frac{1}{2}$, which captures the global Poincar\'{e} translations within the family of supertranslations. We see that the powers of $z,\bar{z}$ on the right hand side of the commutator vanish and therefore the action of $P_{-\frac{1}{2},-\frac{1}{2}}$ is just to shift the weights of the primary by $(h,\bar{h})\rightarrow \left(h+\frac{1}{2},\bar{h}+\frac{1}{2}\right)$. This transformation has been previously considered in multiple contexts~\cite{Stieberger:2018onx,Pate:2019lpp,Fotopoulos:2020bqj,Himwich:2021dau} and is sometimes referred to as a weight-shifting operator since its effect is to shift $\Delta\rightarrow \Delta+1$. Its corresponding action on the modes of primary operators is calculated in appendix~\ref{app:translations}. 

In celestial CFTs, this weight-shifting operator is important because it sends Mellin primaries to Mellin primaries\footnote{The action of a translation on light and shadow primaries is different and does not result in a primary field.}. One can see this by taking \eqref{eq:mellin} and noting that the action of $P_{-\frac{1}{2},-\frac{1}{2}}$ is to shift the power in the Mellin transform i.e shifting the conformal dimension. In standard 2D CFT, one does not expect this to be the case since, usually, translations result in descendants of primary fields. Therefore, this is just one of a large set of properties that distinguishes celestial CFTs from conventional 2D CFTs.

The generators $P_{m,n}$ are themselves modes of a $\left(\frac{3}{2},\frac{3}{2}\right)$ primary operator, the supertranslation current, and transform under global conformal symmetry as 
\begin{equation}
\left[L_k,P_{m,n}\right] = \left(\frac{k}{2}-m\right)P_{m+k,n}, \ \ \left[\bar{L}_k,P_{m,n}\right] = \left(\frac{k}{2}-n\right)P_{m,n+k}.
\end{equation}
The current $P(z,\bar{z})$ that these are the modes of is understood to be related to the leading soft graviton mode and is therefore labeled as the generator of supertranslations: the leading asymptotic symmetry of a gravitational theory in asymptotically flat spacetime. Therefore, the commutators above tell us that under a supertranslation the global conformal generators undergo a shift. Namely
\begin{equation} \label{2.5}
L_k' = L_k + \left(\frac{k}{2}-m\right)P_{m+k,n}
\end{equation}
and likewise for the barred generators.\footnote{If one writes the shift of the generator in the usual way by exponentiating $P$ on either side and then using the Campbell-Baker-Hausdorff formula, we see that the series truncates since the $P$'s commute with each other.} It is easy to show that these shifted generators satisfy the $SL(2,\mathbb{C})$ algebra. This is a realization, in terms of generators, of the fact that there are an infinite number of distinct copies of $SL(2,\mathbb{C})$ on the boundary of asymptotically flat spacetime that are related by supertranslations -- the famous problem of angular momentum in general relativity~\cite{Strominger:2017zoo,penrose1982some}. 

\subsection{Parallels with $w$-algebra}
This was packaged beautifully with the introduction of the $w$-algebra~\cite{Guevara:2021abz, Strominger:2021mtt}. Specifically, if we let $H_{h,\bar{h}}$ be a positive-helicity soft-graviton ($J=2$) operator, taking the weights to be $(h,\bar{h}) = (3-q,1-q)$ and light transforming gives the following operator
\begin{equation}
w^q(z,\bar{z}) = \kappa(q)\int_{\mathbb{R}}\frac{d\bar{w}}{2\pi i}\frac{1}{(\bar{z}-\bar{w})^{2q}}H_{3-q,1-q}(z,\bar{w})
\end{equation}
where $\kappa(q)$ is a normalization that can be found in~\cite{Himwich:2021dau}. These operators are defined for $q=1,\frac{3}{2},2,\cdots$. They transform like a primary with weights $(h,\bar{h}) = \left(3-q,q\right)$ which means that they are all dimension $\Delta=3$ operators with varying spin. Decomposing the operators into modes and letting $\hat{w}^q_n \equiv w^q_{q-2,n}$ it has been shown that they satisfy the following algebra
\begin{equation}
\left[\hat{w}_m^p,\hat{w}_n^q\right] = \left[m(q-1)-n(p-1)\right]\hat{w}_{m+n}^{p+q-2}.
\end{equation}
The modes come with the restriction that $1-q\leq n\leq q-1$ which is why this is often referred to as the wedge sub-algebra of $w_{1+\infty}$. 

When $q=1$, the graviton we begin with has $\Delta=2$. This mode generally vanishes and does not correspond to a soft mode~\cite{Guevara:2021abz} but it is necessary to include it here to ensure closure of the algebra. When $q=\frac{3}{2}$, the graviton has $\Delta=1$ which is the leading soft mode. Therefore, this is the supertranslation generator. We see that it has $h=\bar{h} = \frac{3}{2}$ and is precisely the operator whose modes are the $P_{m,n}$ we discussed in the previous subsection. When $q=2$, we have the subleading $\Delta=0$ mode which corresponds to superrotations. The modes $\{\hat{w}^{2}_{\pm 1},\hat{w}^{2}_{0}\}$ of this operator are the three generators of global conformal symmetry.\footnote{Interestingly, there is an additional closed $SL_{2}$ sub-algebra generated by the modes $\{\hat{w}^{\frac{3}{2}}_{-\frac{1}{2}},\hat{w}^{2}_{0},\hat{w}^{\frac{5}{2}}_{\frac{1}{2}}\}$, which respectively correspond to modes of the leading, subleading and sub-subleading soft graviton. Starting from the $\hat{w}^{\frac{3}{2}}_{\frac{1}{2}}$ mode, one can generate any other mode in the $w$-algebra from combined actions of both these distinct $SL_{2}$ sub-algebras~\cite{Banerjee:2023jne}.}

Comparing to the previous discussion, we see that the statements about the transformation of the global conformal generators under supertranslations as well as the action of supertranslations on other operators are contained in the $w$-algebra and its properties. It is also important to note that while the spacetime translations shift the dimension of a Mellin primary operator, since the $w$-algebra generators are light transforms of primaries, translations act on them differently. This action is computed in general in appendix~\ref{app:translations}. In particular, translations shift us within the $w$-algebra and take generators to other generators rather than shifting us out of the wedge.

\section{Corners of celestial diamonds}\label{sec:celdiam}
The symmetries discussed in the previous section are elucidated in terms of celestial operators using the diamond structure introduced in~\cite{Pasterski:2021fjn,Pasterski:2021dqe}. In this section we will review these results using our notation which will be necessary for defining marginal operators in the following section. The insight that led to this construction was the fact that in celestial CFTs some descendants of primaries are also primaries and are thus referred to as primary descendants. Usually descendants are given by derivatives acting on the primary field~\cite{DiFrancesco:639405}. We will first show that for specific primaries in celestial CFT, the action of derivatives can be identified as light and shadow transforms.

Consider the definition of the holomorphic light transform 
\begin{equation}
\mathcal{L}\left[\mathcal{O}_{h,\bar{h}}\right](z,\bar{z}) = \int_{\mathbb{R}}\frac{dw}{2\pi i}\frac{1}{(w-z)^{2-2h}}\mathcal{O}_{h,\bar{h}}(w,\bar{z}).
\end{equation}
Using the identity
\begin{equation}
\frac{1}{\Gamma(n+1)}\partial_y^n\frac{1}{x-y}  = \frac{1}{(x-y)^{n+1}}
\end{equation}
we can rewrite the light transform of the operator as 
\begin{equation}
\mathcal{L}\left[\mathcal{O}_{h,\bar{h}}\right](z,\bar{z}) = \int_{\mathbb{R}}\frac{dw}{2\pi i}\frac{1}{(w-z)^{2-2h}}\mathcal{O}_{h,\bar{h}}(w,\bar{z}) = \frac{1}{\Gamma(2-2h)}\partial_z^{1-2h}\int_{\mathbb{R}}\frac{dw}{2\pi i}\frac{1}{w-z}\mathcal{O}_{h,\bar{h}}(w,\bar{z}).
\end{equation}
Analytically continuing the integral over the real line into a contour integral, we see that it evaluates to the residue
\begin{equation}
\mathcal{L}\left[\mathcal{O}_{h,\bar{h}}\right](z,\bar{z}) = \frac{1}{\Gamma(2-2h)}\partial_z^{1-2h}\mathcal{O}_{h,\bar{h}}(z,\bar{z}).
\end{equation}
Likewise, we can convince ourselves that the anti-holomorphic light transform and shadow transform can be written as 
\begin{eqnarray}
\bar{\mathcal{L}}\left[\mathcal{O}_{h,\bar{h}}\right](z,\bar{z}) & = &  \frac{1}{\Gamma(2-2\bar{h})}\partial_{\bar{z}}^{1-2\bar{h}}\mathcal{O}_{h,\bar{h}}(z,\bar{z})\cr
\mathcal{S}\left[\mathcal{O}_{h,\bar{h}}\right](z,\bar{z}) & = & \frac{1}{\Gamma(2-2h)\Gamma(2-2\bar{h})}\partial_z^{1-2h}\partial_{\bar{z}}^{1-2\bar{h}}\mathcal{O}_{h,\bar{h}}(z,\bar{z}).
\end{eqnarray}
This relationship becomes more realizable if we choose the weights of the initial operator to be $(h,\bar{h}) = \left(\frac{1-k}{2},\frac{1-\bar{k}}{2}\right)$ and call that operator $\mathcal{O}_{k,\bar{k}}(z,\bar{z})$. In this case we see that 
\begin{equation}
\mathcal{L}\left[\mathcal{O}_{k,\bar{k}}\right](z,\bar{z}) = \frac{\partial_z^k\mathcal{O}_{k,\bar{k}}(z,\bar{z})}{\Gamma(1+k)}, \  \bar{\mathcal{L}}\left[\mathcal{O}_{k,\bar{k}}\right](z,\bar{z}) = \frac{\partial_{\bar{z}}^{\bar{k}}\mathcal{O}_{k,\bar{k}}(z,\bar{z})}{\Gamma(1+\bar{k})}, \  \mathcal{S}[\mathcal{O}_{k,\bar{k}}](z,\bar{z}) = \frac{\partial_z^k\partial_{\bar{z}}^{\bar{k}}\mathcal{O}_{k,\bar{k}}(z,\bar{z})}{\Gamma(1+k)\Gamma(1+\bar{k})}.
\end{equation}
We see that these three operators combined with the initial operator, form the four corners of the celestial diamonds discussed in~\cite{Pasterski:2021fjn,Pasterski:2021dqe}. While derivatives of primaries are usually descendants that do not transform as primaries, the authors of~\cite{Pasterski:2021fjn} note that for celestial CFTs these three transformations of a primary operator correspond to primary descendants, meaning that they transform as primaries in the CFT.\footnote{Writing out the Lorentz generators in the bulk four dimensional spacetime and acting on the original operator as well as its light and shadow transformations can be shown to give the action of the conformal generators on the sphere which shows that they transform like primaries in the celestial CFT.} 

\begin{figure}[htb!]
\begin{center}
\includegraphics[scale=.55]{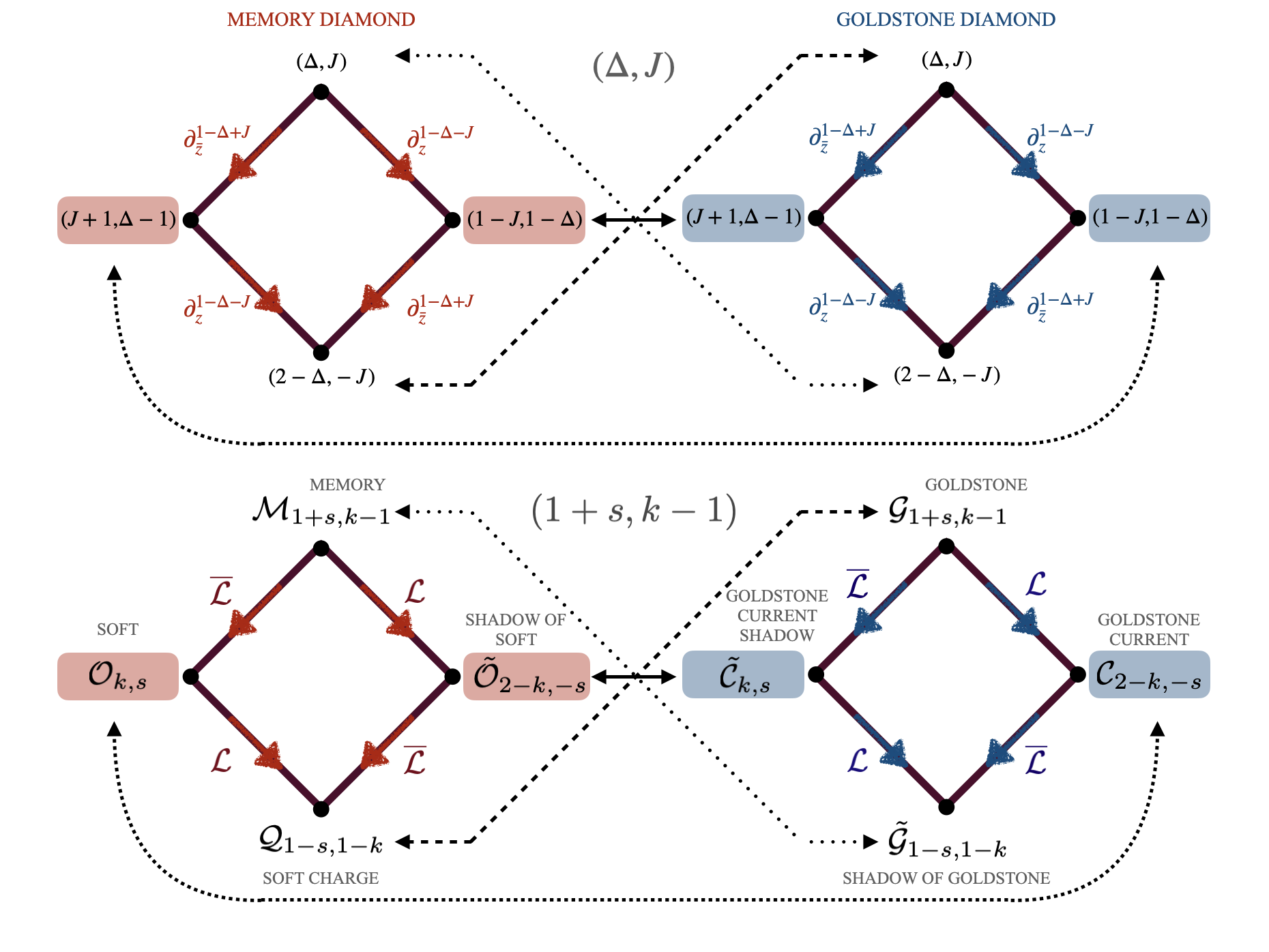}
\caption{We show the pair of diamonds that we are interested in for general labels of spin and dimension. The pair at the top shows the two diamonds for arbitrary $(\Delta,J)$. On the left is the memory diamond, labeled in red, and on the right is the Goldstone diamond, labeled in blue. The corners of the red and blue diamonds are symplectically paired, notated by the arrows. The bottom set of diamonds is for a choice where $(\Delta,J) = (1+s,k-1)$ for the operator at the top corner. In this case, one corner of the memory diamond can be identified as a soft operator of spin $s$ and dimension $k$. The other corners of the diamonds are notated to show their relation to one another as well as to identify them within the operator spectrum of the celestial CFT.}
\label{Fig:diamond}
\end{center}
\end{figure}
The relationship between the operators as defined can be seen in Figure~\ref{Fig:diamond}. The red diamonds are the memory diamonds and the blue are the Goldstone diamonds. Focusing on the bottom set of diamonds we can see that if we make the choice for the top corner operator to have $(\Delta=1+s,J=k-1)$ then we can identify the the side corners of the diamonds with operators that form the integer basis spectrum in celestial CFT~\cite{Freidel:2022skz,Cotler:2023qwh}. We begin with the tops of the diamonds which define their classification since the top of the memory diamond is the memory mode and the top of the Goldstone diamond is the Goldstone mode. The Goldstone mode is the operator that shifts under a given asymptotic symmetry while the memory mode describes the change in a boundary field under the same symmetry. More than these operators on their own, of particular interest are their primary descendants. One primary descendant of the memory mode, which lies at the left corner of the red diamond, can be identified as the soft operator which is the operator that we use to construct the current whose insertion yields the conformally soft theorems. Descending one more level gives us the soft charge which is the conserved charge associated with that asymptotic symmetry. The opposite corner of the memory diamond is the shadow of the soft operator which is also a primary field in the spectrum of the celestial CFT but only in certain cases plays a special role in the soft story. We will discuss this in more detail later.

The powers of derivatives that transport us from corner to corner give us restrictions on the number of diamonds we can have of this form for radiative modes of a given spin, and are equivalently notated as light and shadow transforms as we have shown above.\footnote{This should not be taken to imply that there are no diamond structures for primaries of other weights. Rather, there are three types of primary descendants as discussed in~\cite{Pasterski:2021fjn,Pasterski:2021dqe} and we have chosen to draw the cases where there is a full diamond structure and not a truncated diamond.} Since the powers of the derivatives have to be non-zero and positive, we see that $\Delta-1< J<1-\Delta$ and that $\Delta<1$. We will enumerate what this means for a few relevant cases.
\begin{figure}[t]
\begin{center}
\includegraphics[scale=.7]{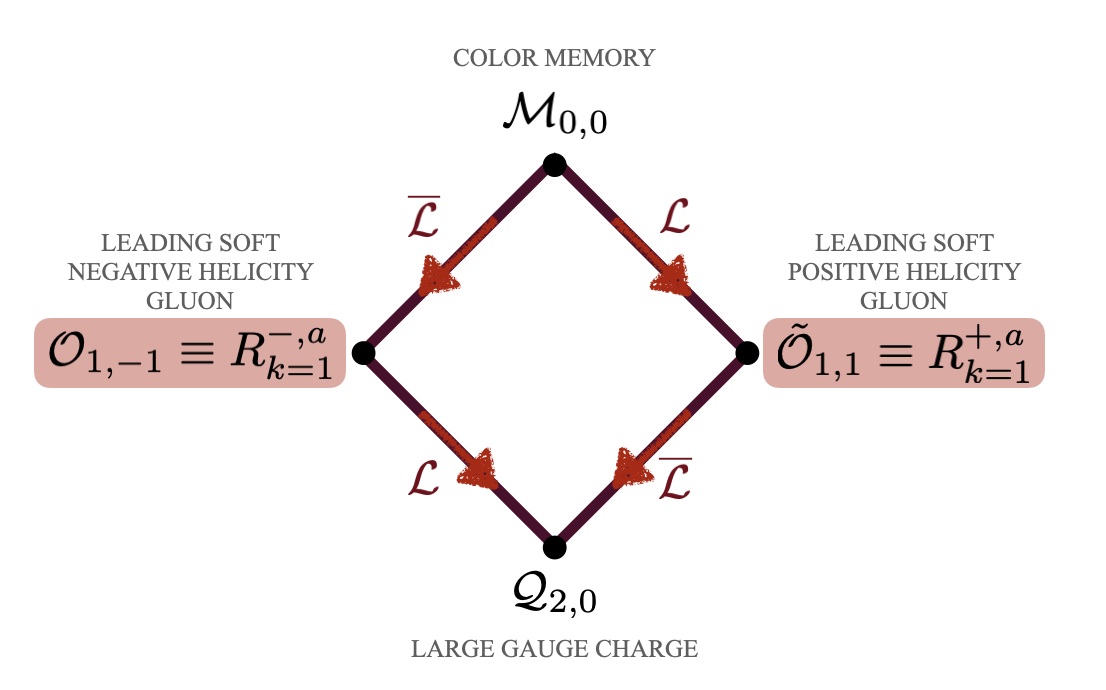}
\caption{The memory diamond in the case where the soft operators are spin-1. This corresponds to the $s=-1$ case in the bottom figure of \ref{Fig:diamond}. $\Delta=1$ for both of the soft operators, we see that they correspond to the same charge, the large gauge charge. A copy of this diamond also exists for photons.}
\label{Fig:spin1dia}
\end{center}
\end{figure}
\paragraph{Scalars} We can start with the simplest case where the soft operators are scalars. Then the parent primary must have $\Delta=1$. In that case, there are no allowed values of $J$ that will give us a diamond that has four points. Rather we will find that all the scalar primaries appear at relevant corners of the diamonds for higher spin soft particles.

\paragraph{Spin one} When the soft operators are spin-$1$, like in the case for photons or gluons, we need the top corner to have $\Delta=0$. In that case, there is exactly one choice for the 2D spin, $J=0$. The associated memory diamond is shown in Figure~\ref{Fig:spin1dia}. We see that the operator at the bottom of this diamond can be identified as a marginal operator and it comes from light transforming either of the soft modes (or shadowing the top mode). This is consistent with the discussion of marginal operators from soft gluons in~\cite{Narayanan:2024qgb}. We also see that the memory and the charge in this case are scalars, which is one instance where the scalar primaries appear since they do not have their own consistent diamond structure.

\begin{figure}[htb]
\begin{center}
\includegraphics[scale=.5]{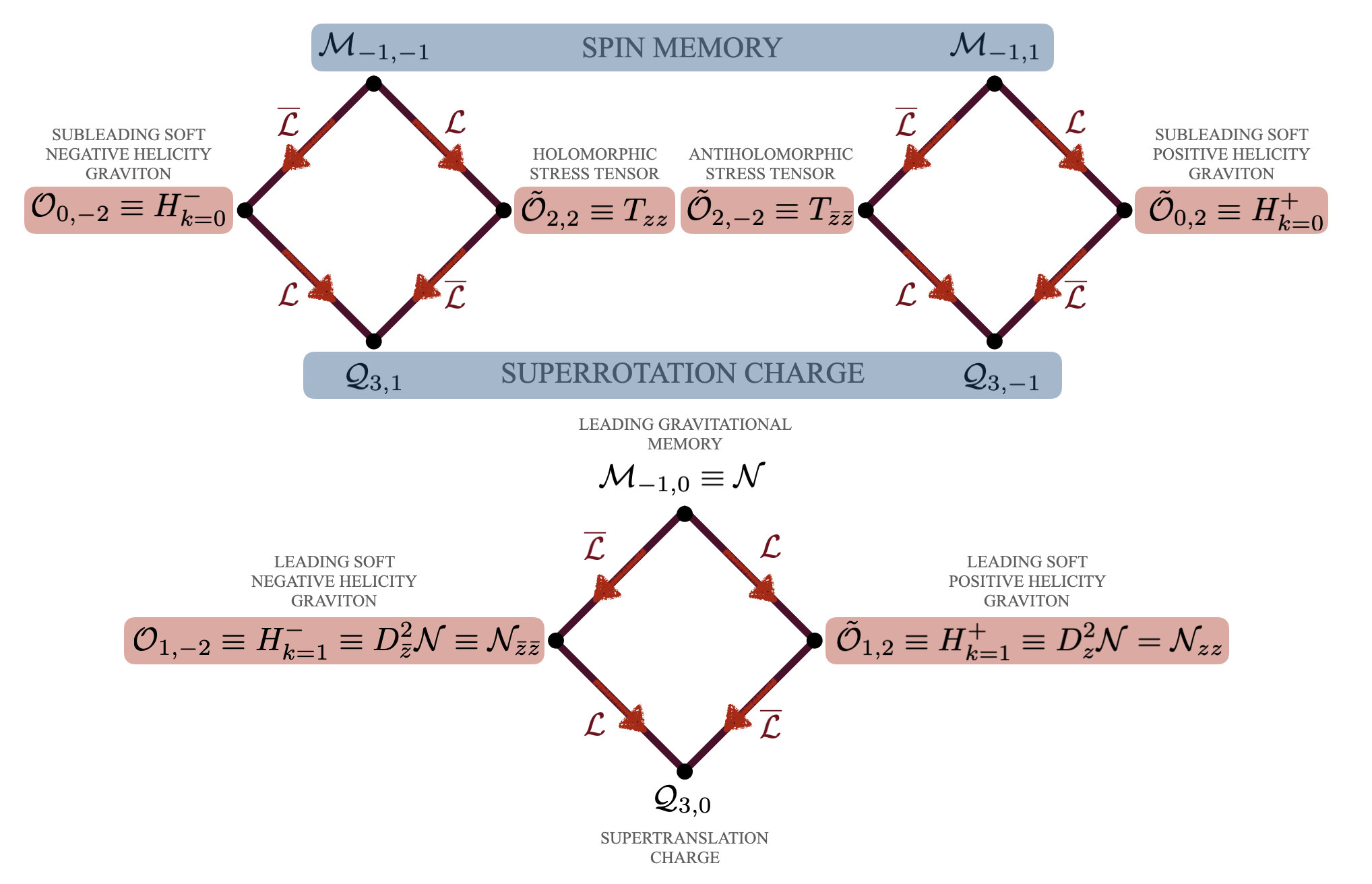}
\caption{The three distinct memory diamonds when we choose the soft operators to be restricted to gravitons. Once again when $\Delta=1$ both soft operators correspond to the same charge as we found in the gauge theory case. However, when $\Delta=0$ each helicity soft operator now has their separate diamond.}
\label{Fig:spin2dia}
\end{center}
\end{figure}

\paragraph{Spin two} When the soft operators are gravitons then the top corner has to have $\Delta=-1$ which means that there are three possible diamonds $J=-1,0,1$ as shown in Figure~\ref{Fig:spin2dia}.
Similar to the gauge theory case, when the soft operators have $\Delta=1$, the opposite helicity operators appear in the same diamond and correspond to the same soft charge since they are shadows of each other. In the gravity case, the subleading soft operator is also necessary for the symmetries on the boundary and hence also has its own diamond. However, since the opposite helicity operators are not shadows of each other in this case, they do not correspond to the same charge at the bottom of the diamond and additionally originate from memory operators of different helicities as can be seen from the two top diamonds in Figure~\ref{Fig:spin2dia}. We also see that the three graviton diamonds contain new scalar primaries as well as spin-1 primaries that did not have their own diamond structure. 

In this way, we can see a pattern emerging for higher spin diamonds. If we consider soft operators that are spin-$J$, the memory operator (or Goldstone in the conjugate diamond) will need to have $\Delta = 1-|J|$ so there will be $2|J|-1$ diamonds. The rest of the relevant primaries will appear as corners in other higher spin diamonds. For each $J$, there will always be a single diamond whose memory and charge are scalars and the other two corners are soft operators that degenerate because their dimension is $\Delta=1$. One reason why this happens is because when $\Delta=1$, the operator at the top of the diamond is a scalar. The copies of diamonds come from needing to consider different helicities as in the gravitational case for $\Delta=0$.

Lastly, it is important to comment on how translations, as defined in section~\ref{sec:celsym} act within the diamonds. Rather than taking primaries to primary descendants, a single translation takes the operator to an operator that lies on an edge of the diamond rather than a corner. Therefore, it is not possible to begin with a primary operator and reach a primary descendant via the use of translations. This is not surprising, since we know that $\partial_z,\partial_{\bar{z}}$ will shift $h,\bar{h}$ by 1, not by $\frac{1}{2}$ which means that a single translation will not be equivalent to a set of derivatives acting on a primary. We have emphasized the importance of the corners of celestial diamonds since they will be necessary in the construction of a universal set of marginal operators in the next section. 

\section{Marginal operators in gravity}\label{sec:marginal}
Now that we have an understanding of the symmetries and spectrum of celestial CFTs, we are ready to discuss the construction of marginal operators. Given a CFT, one can deform it by the addition of an operator into the spectrum that falls into one of three categories: relevant, irrelevant or marginal. In 2D, a relevant operator is one that has conformal dimension $\Delta<2$ while an irrelevant operator has $\Delta>2$. The addition of either of these types of operators into the spectrum of a CFT will break the conformal symmetry and result in a theory that is no longer conformally invariant. Marginal operators, those with $\Delta=2$, are special because adding them to the spectrum preserves the conformal symmetry and therefore gives rise to a separate but equally well-defined CFT. 

The space of all such deformations is known in the literature as a conformal manifold. Each point on the manifold is understood to be a distinct CFT and one moves around in the manifold with directions specified by the marginal operators. The geometry of this manifold is nicely encoded in the operator product expansions of the marginal operators. Namely, in~\cite{Kutasov:1988xb} it is given by
\begin{equation}\label{eq:margOPE}
\mathbb{M}_I(x)\mathbb{M}_J(y)\sim \frac{g_{IJ}}{|x-y|^4} + \Gamma_{IJ}^K \mathbb{M}_K(y)\delta^{(2)}(x-y)+\cdots
\end{equation}
where the extra terms are regular. We see that the leading term in the OPE is proportional to the metric on the conformal manifold and the next to leading term, that is proportional to a delta function, gives the connections. The regular terms are related to the curvature tensor and its derivatives. 

In the context of celestial holography, exploring the space of marginal deformations provides a potential description for the entire space of possible boundary CFTs. An understanding of this space could help improve what we know about the corresponding bulk theories. Initially in~\cite{Kapec:2022axw}, the aforementioned manifold was studied in the context of a bulk non-linear sigma model (NLSM) which meant that the fundamental fields were scalar fields. The advantage to scalar fields is that their associated boundary operators already have spin-$0$ and therefore, one just needs to choose the conformal dimension to be $\Delta=2$ to get the entire space of marginal operators.

For bulk theories of higher spin particles, like in gauge theory and gravity, the construction of marginal operators is more complicated. As mentioned in the introduction, the preservation of the Casimir allows a natural construction of marginal operators from single soft particles in the scalar and spin-$1$ cases but not for higher spin. The conclusions we have reached simply by looking at the Casimirs are also consistent with statements that come from considering the celestial diamonds, as in the previous section. In particular, we note that the primary descendants in the spin-1 case are precisely the ones found in~\cite{Narayanan:2024qgb} which means, naively, there are no other single particle contenders for marginal operators in either of these cases. This therefore begs the question of how we can construct marginal operators in gravity. 

\subsection{Single particle operators}

If we insist on restricting ourselves to single particle operators then one natural option is to construct marginal operators that are descendants. These operators will take the general form
\begin{equation}
\mathbb{M}^{n,m}(z,\bar{z}) = \partial_z^n\partial_{\bar{z}}^m \mathcal{O}_{h,\bar{h}}(z,\bar{z})
\end{equation}
where $\mathcal{O}_{h,\bar{h}}(z,\bar{z})$ is a primary operator. Since a primary operator admits a mode expansion of the following form 
\begin{equation}
\mathcal{O}^{h,\bar{h}}(z,\bar{z}) = \sum_{(p,q)\in\mathbb{Z}-(h,\bar{h})}\frac{\mathcal{O}^{h,\bar{h}}_{p,q}}{z^{p+h}\bar{z}^{q+\bar{h}}}
\end{equation}
we see that the operator $\mathbb{M}^{n,m}$ has weights $(h+n, \bar{h}+m)$. In order for this operator to be marginal, we need $(n,m) = (1-h, 1-\bar{h})$ or, equivalently, $(h,\bar{h})=(1-n,1-m)$. Since both $n$ and $m$ have to be non-negative integers, $(h,\bar{h})$ must be integers themselves. If we consider the tower of conformally soft graviton modes
\begin{equation}
    H^{k}_+(z,\bar{z})=\lim_{\varepsilon\to 0}\varepsilon\mathcal{O}^{+}_{k+\varepsilon}(z,\bar{z})
\end{equation}
which have dimension $(h,\bar{h})=(\frac{k+2}{2},\frac{k-2}{2})$ with $k=2,1,0,-1,-2,\cdots$, marginal descendants must obey $(\frac{k+2}{2},\frac{k-2}{2})=(1-n,1-m)\in\mathbb{Z}$. This forces $k$ to be an \textit{even} integer. Therefore, we are restricted to the modes $k=2,0,-2,-4,\cdots$. 

Notice that this implies one cannot construct a marginal descendant starting from a leading soft mode, which is the $k=1$ mode and generator of supertranslations. However, one can construct a marginal descendant from a subleading soft mode ($k=0$), which is the generator of Virasoro superrotations. This operator will correspond to $(n,m)=(0,2)$ i.e.
\begin{equation}
    \mathbb{M}^{(0,2)}(z,\bar{z})=\partial_{\bar{z}}^{2}H_+^{0}(z,\bar{z})
\end{equation}
One can easily verify that this operator has the correct dimensions to be exactly marginal, i.e. has weights $(h,\bar{h})=(1,1)$. However, in order to perform marginal deformations of a CFT, one needs to have exactly marginal \textit{primary} operators. Since these operators do not have the correct number of derivatives to be identified as a primary descendants, they are not primaries. Therefore, the marginal descendants, as defined above, cannot generate deformations to the celestial CFT. 

An alternative proposition is to utilize the generator of spacetime translations which takes Mellin primaries to other Mellin primaries in celestial CFT. As discussed in section~\ref{sec:celsym}, the generator $P_{-\frac{1}{2},-\frac{1}{2}}$ shifts the conformal dimension $\Delta$ by 1. It is reasonable to think that combinations of the translations with shadow and light transforms might allow us to construct single particle marginal operators that are primaries. If we start with a soft graviton $H^k_+$, the action of a translation will shift us to $H^{k+1}_+$. We already know that there exists no $k$ such that the light transform or shadow transform of $H^k_+$ is marginal, which means we will not be able to obtain a single particle marginal primary by shifting a graviton and then doing an integral transform. Therefore, the only possibility is to take a light or shadow transform of a soft graviton and then act with $P_{-\frac{1}{2},-\frac{1}{2}}$ on that operator. We show in appendix~\ref{app:translations} that acting with a translation on a light or shadowed primary, does not result in a primary operator. This can be phrased as the fact that the integral transforms do not commute with translations. Therefore, these will not be contenders for marginal deformations either. This should not be surprising since as discussed in~\cite{Pasterski:2021fjn}, acting on any of the corners in the celestial diamond with a translation shifts to a point between the corners which is not a primary descendant.

This further solidifies the statement that the only relevant single particle marginal operators that can be constructed are those that are primary descendants and appear on the corners of the celestial diamonds in gauge theory but not in gravity. If we want to discuss operators that can actually generate deformations in celestial CFTs and have an interpretation in terms of the conformal manifold, then we need to consider composite operators. If we label $H_{h,\bar{h}}$ as a soft graviton operator with spin $J=\pm 2$ and dimension $\Delta$ then its shadow $\tilde{H}_{1-h,1-\bar{h}}$ is an operator with opposite spin and dimension $2-\Delta$. The composite operator made of this pair will therefore be a scalar of dimension $\Delta'=\Delta+(2-\Delta)=2$. Such a composite operator can therefore generate a marginal deformation. In fact there is a set of such composite operators that can be constructed for celestial CFTs which we explore in the following section.

\subsection{General construction}
The issue outlined above tells us that only in the case of scalars and spin-1 particles, we are able to construct marginal operators from single soft operators. This is also evident from the discussion of celestial diamonds since these operators appear within the diamonds. In this subsection we propose a construction for marginal operators starting with any spin. We are given the freedom to work with any basis of operators that span the celestial CFT so we use the set of soft operators $\mathcal{O}_{k,s}$ to be the starting point. Using the notation in Figure~\ref{Fig:diamond}, the shadow of this operator is $\tilde{\mathcal{O}}_{2-k,-s}$ and the symplectic partners of both of these are the Goldstone current $\mathcal{C}_{2-k,-s}$ and its shadow $\tilde{\mathcal{C}}_{k,s}$. Given these four distinct operators, which we can think of as being in a ``family" generated by the initial operator, we can construct the following set of multi-particle operators 
\begin{eqnarray}
\mathbb{M}_{\mathcal{O}\tilde{\mathcal{O}}}(z,\bar{z}) \equiv  :\mathcal{O}_{k,s}\tilde{\mathcal{O}}_{2-k,-s}:(z,\bar{z}), & \ \ &\mathbb{M}_{\mathcal{O}\mathcal{C}}(z,\bar{z})  \equiv  :\mathcal{O}_{k,s}\mathcal{C}_{2-k,-s}:(z,\bar{z})\cr
\mathbb{M}_{\tilde{\mathcal{O}}\tilde{\mathcal{C}}}(z,\bar{z}) \equiv  :\tilde{\mathcal{O}}_{2-k,-s}\tilde{\mathcal{C}}_{k,s}:(z,\bar{z}), & \ \ & \mathbb{M}_{\mathcal{C}\tilde{\mathcal{C}}}(z,\bar{z})  \equiv  :\mathcal{C}_{2-k,-s}\tilde{\mathcal{C}}_{k,s}:(z,\bar{z})
\end{eqnarray}
which all have conformal weights $(h,\bar{h})_{\mathbb{M}}  = (1,1)$. If we consider two primaries, as shown in appendix~\ref{app:compprim}, their corresponding composite operator will also be a primary as long as the OPE between them is non-singular.\footnote{We assume that for celestial CFT a primary along with its light and shadow transforms all obey the traditional OPE with the stress tensor.}

All four of these operators are candidates to generate marginal deformations of the celestial CFT for every spin which means that we no longer have the issue that we ran into for gravity. Ideally we would like to show that these operators make sense as marginal operators by computing their OPEs. Rather than the more complex case of gravity, we stick with a simpler example and see what happens in gauge theory. If we let $J=1$ and consider the composite operator consisting of the $\Delta=1$ primary and its shadow, we can see as computed in appendix~\ref{app:compOPE} that the OPE is 
\begin{equation} \label{4.7}
\mathbb{M}(z,\bar{z})\mathbb{M}(w,\bar{w}) \sim \frac{2Nk^2}{|z-w|^4}+\frac{N}{2\pi^2}\delta^{(2)}(w-z)\mathbb{M}(w,\bar{w})+\cdots
\end{equation}
where the $\cdots$ denote regular terms. We expect a similar relation to hold in the case of gravity. In this case, the the composite operator consisting of $J,\bar{J}$ is a primary because the $J\bar{J}$ OPE is non-singular. This must be the case in order to satisfy the Jacobi identity~\cite{Himwich:2025bza}. The OPE \eqref{4.7} is interesting in its own right since the leading term encodes geometric information about the conformal manifold of the celestial CFT dual to bulk Yang-Mills. Indeed, assuming a non-trivial level $k$ for the Kac-Moody algebra satisfied by leading soft gluon currents, \eqref{4.7} indicates that it is likely this conformal manifold is endowed with a non-trivial metric that depends on $k$ and $N$, consistent with the findings of~\cite{Narayanan:2024qgb}.  

In general, not all four of these multi-particle operators will be primaries and therefore, will not all be able to generate marginal deformations. In particular, since an operator and its symplectic partner have a delta function inner product, it is expected that their OPE will be regular whereas an operator and its shadow are generally expected to have a singular OPE. Therefore, it is likely that the only primary marginal operators could be $\mathbb{M}_{\mathcal{O}\tilde{\mathcal{C}}}$ and $\mathbb{M}_{\tilde{\mathcal{O}}\mathcal{C}}$. We will see in the next section that this is consistent with prior statements about marginal deformations and vacuum transitions in gravity and gauge theory. 

This might, however, seem in contradiction with the example above in gauge theory since we used the current $J$ and its shadow $\bar{J}$. This particular example is special since it is for the case when $\Delta=1$. For $\Delta=1$, the soft operator and its shadow degenerate. This was discussed in detail in~\cite{Donnay:2018neh,Donnay:2020guq} and it was shown that in order to construct the symplectic partners of the Goldstone modes, one needed to introduce a log term. If that log term is not present then it appears that the conformally soft mode and the Goldstone mode are indistinguishable. Therefore, due to this ambiguity for $\Delta=1$, we are able to consider the current $J$ and its shadow as equivalent to the current and its symplectic partner. Once we move away from $\Delta=1$ soft modes, we no longer have this issue since the shadows do not degenerate.

\section{Vacuum structure and conformal manifold}\label{sec:vacuum}
In this section we argue that the multi-particle marginal operators constructed in the previous section, in particular those built from conformally soft modes and their symplectic Goldstone partners, can be used to deform the celestial CFT. Such marginal deformations will be dual to bulk vacuum transitions just as in AdS/CFT and other holographic dualities. 

The starting point is to notice that the author of~\cite{Kapec:2022hih} constructs deformations of the intrinsic 2D soft effective action that involve a product of a leading conformally soft mode together with a Goldstone-like operator. That is, they perform the following shift
\begin{equation} \label{5.1}
    S_{\rm soft}\rightarrow S_{\rm soft}[\mathcal{C}]=S_{\rm soft} +\tilde{\mathcal{C}}\cdot \mathcal{S}
\end{equation}
where $S_{\rm soft}$ is the 2D soft effective action of~\cite{Kapec:2021eug} and $\tilde{\mathcal{C}}\cdot \mathcal{S}$ stands for $\tilde{\mathcal{C}}^{a}\mathcal{S}_{a}$ in gauge theory or $\tilde{\mathcal{C}}^{ab}\mathcal{S}_{ab}$ in gravity. Here, the operator $\tilde{\mathcal{C}}$ is the shadow of the Goldstone-like operator $\mathcal{C}$ that~\cite{Kapec:2022hih} argues labels a particular bulk vacuum. As such, $S$-matrix elements carry a label $\langle\cdots\rangle_{\mathcal{C}}$ to specify which bulk vacua one is in during a scattering event. Following the structure of celestial diamonds shown in figure~\ref{Fig:diamond}, $\mathcal{S}$ is the leading  conformally soft mode at the left corner of the memory diamond that we have called $\mathcal{O}_{k,s}$ while it's symplectic partner, $\tilde{\mathcal{C}}$, lies at the right corner of the Goldstone diamond which we have denoted as $\mathcal{C}_{2-k,-s}$. At face, this seems like a contradiction since~\cite{Kapec:2022hih} says that the symplectic partner is the shadow of the Goldstone while we say that it is the Goldstone. However, we should keep in mind that the discussion in~\cite{Kapec:2022hih} is for leading soft theorems in gauge theory and gravity, where the soft operators have $\Delta=1$. In that case, the degeneracy that we discussed allows us to have two possible contenders for the deformation to the action: $\tilde{\mathcal{C}}\cdot\mathcal{S}$ and $\mathcal{C}\cdot\tilde{\mathcal{S}}$. In the leading case, these correspond to deformations where the soft operator is of different spin so the choice of either does not change the interpretation.

Taking derivatives of a celestial correlation function with respect to $\tilde{\mathcal{C}}$ inserts conformally soft modes into the $S$-matrix, while turning on a coherent state of soft gravitons induces a vacuum transition. In gravity, these two statements are, respectively, given by ~\cite{Kapec:2022hih}\footnote{For a ($d+2$)-dimensional bulk, the constant $c_{1,2}=\pi^{2}$ when $d=2$. For $d>2$ it has a more complicated formula. See~\cite{Kapec:2021eug} for more details.}

\begin{equation} \label{5.2}
\langle{\mathcal{S}_{ab}X\rangle}_{\mathcal{C}_{ab}} = 16ic_{1,2} \frac{\delta}{\delta\tilde{\mathcal{C}}_{ab}}\langle{X}\rangle_{\mathcal{C}_{ab}}
\end{equation}
\begin{equation} \label{5.3}
    \langle{X\rangle}_{\mathcal{C}_{ab}} = \langle{Xe^{\frac{i}{16c_{1,2}}\int d^{d}x(\tilde{\mathcal{C}}^{ab}\mathcal{S}_{ab})(x)}}\rangle_{\mathcal{C}_{ab}=0}
\end{equation}
where $X$ stands for some string of celestial operators in the correlator. In \eqref{5.3} we have suggestively written $(\tilde{\mathcal{C}}^{ab}\mathcal{S}_{ab})(x)$ in the exponential which simply means that both operators are functions of $x$. Geometrically,~\cite{Kapec:2022hih} argues that \eqref{5.2} is the statement that the insertion of a soft mode $\mathcal{S}$ infinitesimally parallel transports the $S$-matrix about the vacuum manifold in the bulk. Since $\mathcal{S}$ in \eqref{5.1} is a leading ($\Delta=1$) conformally soft mode, the points in this moduli space are related by a leading large gauge transformation in gauge theory or a supertranslation in gravity. As we have already pointed out, each point in this moduli space of vacua comes with its own copy of $SL(2,\mathbb{C})$ which, for the case of gravity, was the statement that \eqref{2.5} satisfies a Lorentz algebra. As~\cite{Kapec:2022hih} points out, this is to be viewed as a feature of asymptotically flat quantum gravity because it gives a solution to the problem of angular momentum in general relativity: 
\begin{center}
\textbf{Different bulk vacua related by supertranslations carry different copies of the Lorentz group because each one contains different numbers of leading soft gravitons. }  
\end{center}

How does this story relate to our construction of exactly marginal operators? Recall from previous sections that a conformally soft mode of spin $J$, $\mathcal{O}_{\Delta,J}(z,\bar{z})$, has $\Delta=1-n$. Its symplectic Goldstone partner, $\mathcal{C}(z,\bar{z})$, has $\Delta=1+n$ and opposite spin.\footnote{As previously mentioned, there is a well-known degeneracy for $n=0$ mode that requires the construction of a logarithmic conformal primary wavefunction~\cite{Donnay:2018neh}. This will not affect our final conclusions however. } The tower of these soft and Goldstone operators for $n\in \mathbb{Z}_{\geq 0}$ form a discrete basis for the spectrum of  celestial CFT~\cite{Freidel:2022skz, Cotler:2023qwh}. Therefore, we can construct the following multi-particle operator
\begin{equation} \label{5.4}
    \mathbb{M}_{n}(z,\bar{z})=:\mathcal{C}^{\mp}_{\Delta=1+n}\mathcal{O}_{\Delta=1-n}^{\pm}:(z,\bar{z})=\oint\frac{dx}{2\pi i}\oint\frac{d\bar{x}}{2\pi i}\frac{1}{x-z}\frac{1}{\bar{x}-\bar{z}}\mathcal{C}^{\mp}(x,\bar{x})\mathcal{O}^{\pm}(z,\bar{z})
\end{equation}
Such an operator is exactly marginal for any value of $n\in\mathbb{Z}_{\geq 0}$, namely for any combination of conformally soft and Goldstone pairs that encode the infinite tower of tree-level soft theorems with arbitrary spin. We claim that the deformation leading to bulk vacuum transitions in~\cite{Kapec:2022hih} is the $n=0$ (leading conformally soft) case and that the combination $\tilde{\mathcal{C}}\cdot\mathcal{S}$ appearing in the deformation \eqref{5.1} should be viewed as an exactly marginal deformation of the intrinsically 2D soft effective action by the multi-particle operator $\mathbb{M}_{0}(z,\bar{z})$. Moreover, just as~\cite{Kapec:2022hih} argued that the bulk $S$-matrix should carry a label $\mathcal{C}$ to specify the bulk vacuum, we argue this should be extended to the entire tower of Goldstone modes with $\Delta=1+n$. In other words, the bulk $S$-matrix should carry an infinite number of labels $\langle{\cdots}\rangle_{(\mathcal{C}_{0},\mathcal{C}_{1},\mathcal{C}_{2},\cdots,\mathcal{C}_{n},\cdots)}$ to specify which (sub)$^{n}$-leading soft vacua a scattering event is taking place in\footnote{Beyond sub-subleading order in the soft expansion, it is not entirely clear whether an IR triangle type structure exists. However, we still claim that there should be a vacuum label attached to the bulk $S$-matrix for every known asymptotic symmetry that generates a soft theorem.}. 

Exactly marginal deformations of the holographic celestial CFT by the multi-particle operator $\mathbb{M}_{n}(z,\bar{z})$ will induce a vacuum transition in the bulk in that it will shift one of the labels that the $S-$matrix carries\footnote{Here, $\lambda_{n}$ is a dimensionless coupling that parametrizes the deformation by the operator $\mathbb{M}_{n}(z,\bar{z})$.}
\begin{equation} \label{5.5}
    S_{celestial CFT}\rightarrow S_{celestial CFT}[\mathcal{C}_{n}]=S_{celestial CFT} + \lambda_{n}\int d^{2}z\mathbb{M}_{n}(z,\bar{z})
\end{equation}
\begin{equation} \label{5.6}
    \langle{X}\rangle_{(\mathcal{C}_{0},\cdots, \mathcal{C}_{n}-\delta\mathcal{C}_{n},\cdots)} = \langle{Xe^{-\lambda_{n}\int d^{2}z\mathbb{M}_{n}(z,\bar{z})}}\rangle_{{(\mathcal{C}_{0},\cdots,\mathcal{C}_{n},\cdots)}}
\end{equation}
for any $n\in\mathbb{Z}_{\geq 0}$. Comparing with \eqref{5.3} we see that the coupling $\lambda_{0}=\frac{1}{16c_{1,2}}$ and the bulk $S$-matrix label $\mathcal{C}$ used in~\cite{Kapec:2022hih} is $\mathcal{C}_{0}$ in \eqref{5.6}. We also see from \eqref{5.6} that there is a one-to-one correspondence between the infinite set of asymptotically flat vacua $(\mathcal{C}_{0},\mathcal{C}_{1},\cdots,\mathcal{C}_{n},\dots)$ and the space of exactly marginal couplings $(\lambda_{0},\lambda_{1},\cdots,\lambda_n,\cdots)$. The former can be viewed as coordinates for the infinite dimensional bulk vacuum manifold, arising from the spontaneous breaking of the (possibly) infinite number of asymptotic symmetries one has. The latter provides coordinates for the conformal manifold, i.e. the space of CFTs connected by exactly marginal deformations. Therefore, there is a natural holographic correspondence between the bulk vacuum manifold and the boundary conformal manifold in celestial holography.

To summarize, we have the following celestial holographic statement: exactly marginal deformations of the celestial CFT by the multi-particle operators $\mathbb{M}_{n}(z,\bar{z})$, where $n\in\mathbb{Z}_{\geq 0}$, induce vacuum transitions in the bulk asymptotically flat spacetime. Equation \eqref{5.6} tells us that this follows from the marginal deformation turning on a coherent state of (sub)$^{n}$-leading soft gravitons, leading to a transition in the corresponding (sub)$^{n}$-leading soft vacuum.

\section{Discussion}\label{sec:future}
We began by outlining the obstacles that make constructing marginal operators in gravity more difficult than in gauge theory. We then showed that there is a special set of composite operators that one can construct that are contenders for marginal operators in celestial CFTs. The fact that they involve a soft operator and its shadow or a soft operator and its symplectic partner, make these operators fit nicely into the picture of celestial diamonds. Additionally, these pairs lend themselves to a bulk interpretation in terms of vacuum transitions. 

We were able to construct marginal operators in four ways using special pairings but it is not necessarily the case that these will always result in deformations. In particular, we mentioned that it is necessary for an operator to be a primary in order to generate a marginal deformation. Since the composite operators are only primaries when the corresponding pairs have a regular OPE, in order to determine whether they generate deformations we need to know the OPE of the primary pairs. While it seems plausible that we could construct other multi-particle operators that have the necessary weights to be marginal, our expectation is that only the ones that have a bulk interpretation in terms of vacuum transitions are operators that generate deformations of the conformal manifold i.e the majority will not be primary operators.

It is known that the antisymmetric double soft limit of leading soft scalars, gluons or gravitons probes the curvature of the corresponding bulk vacuum manifold~\cite{Kapec:2022axw,Kapec:2022hih}. It would be interesting to see whether the marginal operators we have defined are capable of probing the curvature of the subleading soft vacuum, particularly in the gravitational case. This bulk vacuum manifold is expected to have non-trivial curvature since the subleading soft factor does not commute when inserting two subleading soft gravitons into the $S$-matrix. It is likely that this manifests in our work as having a non-trivial OPE between two $\mathbb{M}_{n}(z,\bar{z})$ operators for $n=1$, encoding information about the geometry of the conformal manifold associated to the vacuum label $\mathcal{C}_{1}$.

It would be nice to have a better understanding of why $\Delta=1$ is a special case for higher spin operators. This seems tied to the question of why we are able to construct single particle marginal operators for scalars and spin-$1$ particles but not for higher spin. In particular, it would be satisfying to understand what the interpretation is for the spin-$1$ single particle operators since they do not automatically lend themselves to a description in terms of bulk vacuum states. Additionally, since we now have a rather clean description of the labels for our vacuum state, it would be interesting to see if they have a natural interpretation in terms of a partition function. This might lend itself to a ensemble-like understanding of a quantum field theory with asymptotic symmetries. Some aspects of this have been studied at leading order~\cite{He:2024vlp,He:2025hag} since at subleading order the effective action is not yet understood. It would be interesting to see if our work has any implications for the structure of the soft effective action beyond leading order. We leave this to future work.

\section*{Acknowledgements}
The authors would like to thank Luca Ciambelli, Temple He, Prahar Mitra, Sabrina Pasterski and Monica Pate for insightful discussions. We also thank Sabrina Pasterski for comments on an earlier version of this manuscript. S.N. acknowledges support by the Celestial Holography Initiative at the Perimeter Institute for Theoretical Physics and by the Simons Collaboration on Celestial Holography. M.I. and S.N.'s research at the Perimeter Institute is supported by the Government of Canada through the Department of Innovation, Science and Industry Canada and by the Province of Ontario through the Ministry of Colleges and Universities. The work of A.W.P. is supported by Discovery Grant from the Natural Sciences and Engineering Research Council of Canada. 

\appendix
\section{Action of translations on celestial operators}\label{app:translations}
In this appendix we explicitly write out the action of translations on light transforms and shadow transforms. We thought it was important and potentially useful to others to have this written out explicitly.\footnote{The authors would like to thank Monica Pate for pointing this out.} The integral transforms that are important are the two light transforms and the shadow transform. First define these transformed operators as
\begin{eqnarray}
\mathcal{L}\left[\mathcal{O}_{h,\bar{h}}\right](z,\bar{z}) & = &  \int_{\mathbb{R}}\frac{dw}{2\pi i} \frac{\mathcal{O}_{h,\bar{h}}(w,\bar{z})}{(w-z)^{2-2h}}, \ \ \bar{\mathcal{L}}\left[\mathcal{O}_{h,\bar{h}}\right](z,\bar{z}) = \int_{\mathbb{R}}\frac{d\bar{w}}{2\pi i}\frac{\mathcal{O}_{h,\bar{h}}(z,\bar{w})}{(\bar{w}-\bar{z})^{2-2\bar{h}}}\cr
\widetilde{\mathcal{O}}_{1-h,1-\bar{h}}(z,\bar{z}) & = & \int\frac{d^2w}{(w-z)^{2-2h}(\bar{w}-\bar{z})^{2-2\bar{h}}}\mathcal{O}_{h,\bar{h}}(w,\bar{w}).
\end{eqnarray}
Now we can compute the action of translations on them. For the holomorphic light transform we get
\begin{eqnarray}
\left[P_{-\frac{1}{2},-\frac{1}{2}},\mathcal{L}\left[\mathcal{O}_{h,\bar{h}}\right](z,\bar{z})\right] & = & \int_{\mathbb{R}}\frac{dw}{2\pi i} \frac{\left[P_{-\frac{1}{2},-\frac{1}{2}},\mathcal{O}_{h,\bar{h}}(w,\bar{z})\right]}{(w-z)^{2-2h}} =  \frac{1}{2}\int_{\mathbb{R}}\frac{dw}{2\pi i} \frac{\mathcal{O}_{h+\frac{1}{2},\bar{h}+\frac{1}{2}}(w,\bar{z})}{(w-z)^{2-2(h+\frac{1}{2})+1}}\cr
& = & \frac{\partial_z}{2(1-2h)}\int_{\mathbb{R}}\frac{dw}{2\pi i} \frac{\mathcal{O}_{h+\frac{1}{2},\bar{h}+\frac{1}{2}}(w,\bar{z})}{(w-z)^{2-2(h+\frac{1}{2})}} =  \frac{\partial_z\mathcal{L}\left[\mathcal{O}_{h+\frac{1}{2},\bar{h}+\frac{1}{2}}\right](z,\bar{z})}{2(1-2h)}.
\end{eqnarray}
For the anti-holomorphic light transform
\begin{eqnarray}
\left[P_{-\frac{1}{2},-\frac{1}{2}},\bar{\mathcal{L}}\left[\mathcal{O}_{h,\bar{h}}\right](z,\bar{z})\right] & = & \int_{\mathbb{R}}\frac{d\bar{w}}{2\pi i} \frac{\left[P_{-\frac{1}{2},-\frac{1}{2}},\mathcal{O}_{h,\bar{h}}(z,\bar{w})\right]}{(\bar{w}-\bar{z})^{2-2\bar{h}}} =  \frac{1}{2}\int_{\mathbb{R}}\frac{d\bar{w}}{2\pi i} \frac{\mathcal{O}_{h+\frac{1}{2},\bar{h}+\frac{1}{2}}(z,\bar{w})}{(\bar{w}-\bar{z})^{2-2(\bar{h}+\frac{1}{2})+1}}\cr
& = & \frac{\partial_{\bar{z}}}{2(1-2\bar{h})}\int_{\mathbb{R}}\frac{d\bar{w}}{2\pi i} \frac{\mathcal{O}_{h+\frac{1}{2},\bar{h}+\frac{1}{2}}(z,\bar{w})}{(\bar{w}-\bar{z})^{2-2(\bar{h}+\frac{1}{2})}} =  \frac{\partial_{\bar{z}}\mathcal{L}\left[\mathcal{O}_{h+\frac{1}{2},\bar{h}+\frac{1}{2}}\right](z,\bar{z})}{2(1-2\bar{h})}.
\end{eqnarray}
Finally for the shadow transform
\begin{eqnarray}
\left[P_{-\frac{1}{2},-\frac{1}{2}},\widetilde{\mathcal{O}}_{1-h,1-\bar{h}}(z,\bar{z})\right] & = & \int\frac{d^2w}{(w-z)^{2-2h}(\bar{w}-\bar{z})^{2-2\bar{h}}}\left[P_{-\frac{1}{2},-\frac{1}{2}},\mathcal{O}_{h,\bar{h}}(w,\bar{w})\right]\cr
& = & \frac{1}{2}\int\frac{d^2w}{(w-z)^{2-2(h+\frac{1}{2})+1}(\bar{w}-\bar{z})^{2-2(\bar{h}+\frac{1}{2})+1}}\mathcal{O}_{h+\frac{1}{2},\bar{h}+\frac{1}{2}}(w,\bar{w})\cr
& = & \frac{\partial_z\partial_{\bar{z}}}{2(1-2h)(1-2\bar{h})}\int\frac{d^2w\mathcal{O}_{h+\frac{1}{2},\bar{h}+\frac{1}{2}}(w,\bar{w})}{(w-z)^{2-2(h+\frac{1}{2})}(\bar{w}-\bar{z})^{2-2(\bar{h}+\frac{1}{2})}}\cr
& = & \frac{\partial_z\partial_{\bar{z}}\widetilde{\mathcal{O}}_{\frac{1}{2}-h,\frac{1}{2}-\bar{h}}(z,\bar{z})}{2(1-2h)(1-2\bar{h})}.
\end{eqnarray}
Likewise, we can work out the action of translations on the modes of a generic celestial operator. Let $\mathcal{O}_{h,\bar{h}}(z,\bar{z})$ be a celestial operator. We can write the modes of this operator as
\begin{equation}
\mathcal{O}^{h,\bar{h}}_{k,\ell}  = \frac{1}{(2\pi i)^2}\oint dz z^{k+h-1}\oint d\bar{z}\bar{z}^{\ell+\bar{h}-1}\mathcal{O}_{h,\bar{h}}(z,\bar{z}).
\end{equation}
We would like to compute the action of translations on these modes. We do so as follows
\begin{eqnarray}
\left[P_{m,n},\mathcal{O}^{h,\bar{h}}_{k,\ell}\right] & = & \frac{1}{(2\pi i)^2}\oint dz z^{k+h-1}\oint d\bar{z}\bar{z}^{\ell+\bar{h}-1}\left[P_{m,n},\mathcal{O}_{h,\bar{h}}(z,\bar{z})\right]\cr
& = & \frac{1}{(2\pi i)^2}\oint dz z^{k+h-1}\oint d\bar{z}\bar{z}^{\ell+\bar{h}-1}\frac{1}{2}z^{m+\frac{1}{2}}\bar{z}^{n+\frac{1}{2}}\mathcal{O}_{h+\frac{1}{2},\bar{h}+\frac{1}{2}}(z,\bar{z})\cr
& = & \frac{1}{2}\frac{1}{(2\pi i)^2}\oint dz z^{k+m+\left(h+\frac{1}{2}\right)-1}\oint d\bar{z} \bar{z}^{\ell+n+\left(\bar{h}+\frac{1}{2}\right)-1}\mathcal{O}_{h+\frac{1}{2},\bar{h}+\frac{1}{2}}(z,\bar{z})\cr
& = & \frac{1}{2}\mathcal{O}^{h+\frac{1}{2},\bar{h}+\frac{1}{2}}_{k+m,\ell+n}
\end{eqnarray}
We see that, as expected, translations shift the dimension of the operator.

\section{Composite Primaries}\label{app:compprim}
In this appendix we will do a computation in standard two-dimensional CFT to show the conditions on when a composite of primaries is also a primary operator. Suppose that we let $\mathcal{O}_1,\mathcal{O}_2$ be primary operators and $T$ be the stress tensor. Then we know that 
\begin{equation}
T(z,\bar{z})\mathcal{O}_i(w,\bar{w}) \sim \frac{h_i\mathcal{O}_i(w,\bar{w})}{(z-w)^2} + \frac{\partial\mathcal{O}_i(w,\bar{w})}{z-w}.
\end{equation}
Defining the composite operator as 
\begin{equation}
:\mathcal{O}_1\mathcal{O}_2:(w,\bar{w}) = \oint_w\frac{dx}{x-w}\oint_{\bar{w}}\frac{d\bar{x}}{\bar{x}-\bar{w}}\mathcal{O}_1(x,\bar{x})\mathcal{O}_2(w,\bar{w})
\end{equation}
we should be able to compute the OPE of the stress tensor with this composite operator using Wicks theorem, as in traditional CFT,
\begin{eqnarray}
T(z,\bar{z}):\mathcal{O}_1\mathcal{O}_2:(w,\bar{w}) & \sim & \oint_w\frac{dx}{x-w}\oint_{\bar{w}}\frac{d\bar{x}}{\bar{x}-\bar{w}}T(z,\bar{z})\left(\mathcal{O}_1(x,\bar{x})\mathcal{O}_2(w,\bar{w})\right)\cr
& = & \oint_w\frac{dx}{x-w}\oint_{\bar{w}}\frac{d\bar{x}}{\bar{x}-\bar{w}}\left[\frac{h_1\mathcal{O}_1(x,\bar{x})}{(z-x)^2}+\frac{\partial\mathcal{O}_1(x,\bar{x})}{z-x}\right]\mathcal{O}_2(w,\bar{w})\cr
& + & \oint_w\frac{dx}{x-w}\oint_{\bar{w}}\frac{d\bar{x}}{\bar{x}-\bar{w}}\mathcal{O}_1(x,\bar{x})\left[\frac{h_2\mathcal{O}_2(w,\bar{w})}{(z-w)^2}+\frac{\partial\mathcal{O}_2(w,\bar{w})}{z-w}\right]\cr
& = & h_1\oint_w\frac{dx}{x-w}\oint_{\bar{w}}\frac{d\bar{x}}{\bar{x}-\bar{w}}\frac{1}{(z-x)^2}\left[\mathcal{O}_1(x,\bar{x})\mathcal{O}_2(w,\bar{w})+:\mathcal{O}_1\mathcal{O}_2:(w,\bar{w})\right]\cr
& + &  \oint_w\frac{dx}{x-w}\oint_{\bar{w}}\frac{d\bar{x}}{\bar{x}-\bar{w}}\frac{1}{z-x}\left[\partial\mathcal{O}_1(x,\bar{x})\mathcal{O}_2(w,\bar{w})+:\partial\mathcal{O}_1\mathcal{O}_2:(w,\bar{w})\right]\cr
& + & \frac{h_2}{(z-w)^2}\oint_w\frac{dx}{x-w}\oint_{\bar{w}}\frac{d\bar{x}}{\bar{x}-\bar{w}}\left[\mathcal{O}_1(x,\bar{x})\mathcal{O}_2(w,\bar{w})+:\mathcal{O}_1\mathcal{O}_2:(w,\bar{w})\right]\cr
& + & \frac{1}{z-w}\oint_w\frac{dx}{x-w}\oint_{\bar{w}}\frac{d\bar{x}}{\bar{x}-\bar{w}}\left[\mathcal{O}_1(x,\bar{x})\partial\mathcal{O}_2(w,\bar{w})+:\mathcal{O}_1\partial\mathcal{O}_2:(w,\bar{w})\right]\cr
& = & \oint_w\frac{dx}{x-w}\oint_{\bar{w}}\frac{d\bar{x}}{\bar{x}-\bar{w}}\left[\frac{h_1}{(z-x)^2}+\frac{h_2}{(z-w)^2}\right]\left[\mathcal{O}_1(x,\bar{x})\mathcal{O}_2(w,\bar{w})\right]\cr
& + &  \oint_w\frac{dx}{x-w}\oint_{\bar{w}}\frac{d\bar{x}}{\bar{x}-\bar{w}}\frac{1}{z-x}\partial_x\left[\mathcal{O}_1(x,\bar{x})\mathcal{O}_2(w,\bar{w})\right]\cr
& + & \oint_w\frac{dx}{x-w}\oint_{\bar{w}}\frac{d\bar{x}}{\bar{x}-\bar{w}}\frac{1}{z-w}\partial_w\left[\mathcal{O}_1(x,\bar{x})\mathcal{O}_2(w,\bar{w})\right]\cr
& + & \frac{h_1+h_2}{(z-w)^2}:\mathcal{O}_1\mathcal{O}_2:(w,\bar{w}) + \frac{1}{z-w}\partial:\mathcal{O}_1\mathcal{O}_2:(w,\bar{w})\cr
& + &  \mathbf{\oint_w\frac{dx}{x-w}\oint_{\bar{w}}\frac{d\bar{x}}{\bar{x}-\bar{w}}\frac{x-w}{(z-x)(z-w)}:\partial\mathcal{O}_1\mathcal{O}_2:(w,\bar{w}) }
\end{eqnarray}
The pairs of operators here mean that we need to insert the OPE. The bold term goes to zero since the singularity at $x-w$ is canceled. We see here that if there are no singular terms in the OPE of $\mathcal{O}_1$ and $\mathcal{O}_2$ then all the bracketed terms vanish and the composite operator is a primary. However, if there are singular terms in their OPE, the composite operator is not necessarily a primary unless those terms conveniently cancel for some other special reason.

\section{OPEs of composites}\label{app:compOPE}
In this appendix we will compute the OPE of a composite operator in gauge theory as an example. Leading soft gluons are known to be dual to Kac-Moody currents. Assuming that we have a current algebra with non-trivial level
\begin{eqnarray}
J^a(z_1,\bar{z}_1)J^b(z_2,\bar{z}_2) & \sim & \frac{k\delta^{ab}}{z_{12}^2}+\frac{if^{ab}_cJ^c(z_2,\bar{z}_2)}{z_{12}}\cr
\bar{J}^a(z_1,\bar{z}_1)\bar{J}^b(z_2,\bar{z}_2) & \sim & \frac{k\delta^{ab}}{\bar{z}_{12}^2}+\frac{if^{ab}_c\bar{J}^c(z_2,\bar{z}_2)}{\bar{z}_{12}}\cr
J^a(z_1,\bar{z}_1)\bar{J}^b(z_2,\bar{z}_2) & \sim & 0
\end{eqnarray}
we will define the following operator.
\begin{equation}
\mathcal{O}(z,\bar{z}) = :J^a\bar{J}^a:(z,\bar{z}) \equiv \frac{1}{(2\pi i)^2}\oint_z\frac{dw}{w-z}\oint_{\bar{z}}\frac{d\bar{w}}{\bar{w}-\bar{z}}J^a(w,\bar{w})\bar{J}^a(z,\bar{z})
\end{equation}
We can compute the OPE as follows. First we derive the OPE with a single current. Using generalized Wick's theorem, we have
\begin{eqnarray}
J^a(z,\bar{z})\mathcal{O}(w,\bar{w}) & = & \frac{1}{(2\pi i)^2}\oint_w\frac{dx}{x-w}\oint_{\bar{w}}\frac{d\bar{x}}{\bar{x}-\bar{w}}J^a(z,\bar{z})\left(J^b(x,\bar{x})\bar{J}^b(w,\bar{w})\right)\cr
& = & \frac{1}{(2\pi i)^2}\oint_w\frac{dx}{x-w}\oint_{\bar{w}}\frac{d\bar{x}}{\bar{x}-\bar{w}}\left[\frac{k\delta^{ab}}{(z-x)^2}+\frac{if^{ab}_cJ^c(x,\bar{x})}{z-x}\right]\bar{J}^b(w,\bar{w})\cr
& = & \frac{k\bar{J}^a(w,\bar{w})}{(z-w)^2}+\frac{if^{ab}_c:J^c\bar{J}^b:(w,\bar{w})}{z-w}
\end{eqnarray}
and likewise with the barred current
\begin{eqnarray}
\bar{J}^a(z,\bar{z})\mathcal{O}(w,\bar{w}) & = & \frac{1}{(2\pi i)^2}\oint_w\frac{dx}{x-w}\oint_{\bar{w}}\frac{d\bar{x}}{\bar{x}-\bar{w}}\bar{J}^a(z,\bar{z})\left(J^b(x,\bar{x})\bar{J}^b(w,\bar{w})\right)\cr
& = & \frac{1}{(2\pi i)^2}\oint_w\frac{dx}{x-w}\oint_{\bar{w}}\frac{d\bar{x}}{\bar{x}-\bar{w}}J^b(x,\bar{x})\left[\frac{k\delta^{ab}}{(\bar{z}-\bar{w})^2}+\frac{if^{ab}_c\bar{J}^c(w,\bar{w})}{\bar{z}-\bar{w}}\right]\cr
& = & \frac{kJ^a(w,\bar{w})}{(\bar{z}-\bar{w})^2}+\frac{if^{ab}_c:J^b\bar{J}^c:(w,\bar{w})}{\bar{z}-\bar{w}}.
\end{eqnarray}
Now we can compute the full OPE
\begin{eqnarray}
\mathcal{O}(z,\bar{z})\mathcal{O}(w,\bar{w}) & = & \frac{1}{(2\pi i)^2}\oint_z\frac{dx}{x-z}\oint_{\bar{z}}\frac{d\bar{x}}{\bar{x}-\bar{z}}\left(J^a(x,\bar{x})\bar{J}^a(z,\bar{z})\right)\mathcal{O}(w,\bar{w})\cr
& = & \frac{1}{(2\pi i)^2}\oint_z\frac{dx}{x-z}\oint_{\bar{z}}\frac{d\bar{x}}{\bar{x}-\bar{z}}\left(\frac{k\bar{J}^a(w,\bar{w})}{(x-w)^2}+\frac{if^{ab}_c:J^c\bar{J}^b:(w,\bar{w})}{x-w}\right)\bar{J}^a(z,\bar{z})\cr
& + & \frac{1}{(2\pi i)^2}\oint_z\frac{dx}{x-z}\oint_{\bar{z}}\frac{d\bar{x}}{\bar{x}-\bar{z}}J^a(x,\bar{x})\left(\frac{kJ^a(w,\bar{w})}{(\bar{z}-\bar{w})^2}+\frac{if^{ab}_c:J^b\bar{J}^c:(w,\bar{w})}{\bar{z}-\bar{w}}\right)\cr
& = & \frac{2Nk^2}{(z-w)^2(\bar{z}-\bar{w})^2} + \left(\frac{if^{ab}_c:J^c\bar{J}^b:(w,\bar{w})}{z-w}\right)\bar{J}^a(z,\bar{z})\cr
& + &  J^a(z,\bar{z})\left(\frac{if^{ab}_c:J^b\bar{J}^c:(w,\bar{w})}{\bar{z}-\bar{w}}\right).
\end{eqnarray}
We can compute the latter two terms by expanding the normal ordered pair
\begin{eqnarray}
\frac{if^{ab}_c:J^c\bar{J}^b:(w,\bar{w})}{z-w}\bar{J}^a(z,\bar{z}) & = & \frac{if^{ab}_c}{(2\pi i)^2}\oint_w\frac{dx}{x-w}\oint_{\bar{w}}\frac{d\bar{x}}{\bar{x}-\bar{w}}\frac{\left(J^c(w,\bar{w})\bar{J}^b(x,\bar{x})\right)}{z-x}\bar{J}^a(z,\bar{z})\cr
& = & \frac{if^{ab}_c}{(2\pi i)^2}\oint_w\frac{dx}{x-w}\oint_{\bar{w}}\frac{d\bar{x}}{\bar{x}-\bar{w}}\frac{J^c(w,\bar{w})}{z-x}\left[\frac{k\delta^{ba}}{(\bar{x}-\bar{z})^2}+\frac{if^{ba}_d\bar{J}^d(x,\bar{x})}{\bar{x}-\bar{z}}\right]\cr
& = & -\frac{N}{(2\pi i)^2}\oint_w\frac{dx}{x-w}\oint_{\bar{w}}\frac{d\bar{x}}{\bar{x}-\bar{w}}\frac{J^c(w,\bar{w})}{x-z}\frac{\bar{J}^c(x,\bar{x})}{\bar{x}-\bar{z}}\cr
& = & -\frac{N}{(2\pi i)^2}\oint_w\frac{dx}{x-w}\oint_{\bar{w}}\frac{d\bar{x}}{\bar{x}-\bar{w}}\frac{J^c(w,\bar{w})}{x-z}\frac{1}{\bar{x}-\bar{z}}\cr
& \times & \sum_{n,m=0}^\infty \frac{(x-w)^n(\bar{x}-\bar{w})^m}{n!m!}\left[\partial^n\bar{\partial}^m\bar{J}^c(x,\bar{x})\right]\bigg|_{(x,\bar{x})\rightarrow (w,\bar{w})}\cr
& = & -\frac{N}{(2\pi i)^2}\oint_w\frac{dx}{x-w}\oint_{\bar{w}}\frac{d\bar{x}}{\bar{x}-\bar{w}}\frac{J^c(w,\bar{w})}{x-z}\frac{\bar{J}^c(w,\bar{w})}{\bar{x}-\bar{z}}\cr
& = & -\frac{N}{(2\pi i)^2}\delta^{(2)}(w-z)\mathcal{O}(w,\bar{w})
\end{eqnarray}
and similarly for the other term. Therefore we get 
\begin{equation}
\mathcal{O}(z,\bar{z})\mathcal{O}(w,\bar{w}) \sim \frac{2Nk^2}{|z-w|^4}-\frac{2N}{(2\pi i)^2}\delta^{(2)}(w-z)\mathcal{O}(w,\bar{w})
\end{equation}
where we have used the fact that $f^{ab}_c f^{ab}_d = N\delta_{cd}$ and that $f^{aa}_b=0$. 
\bibliographystyle{utphys}
\bibliography{cpb}

\end{document}